# High sub-bandgap response and fast switching enabled by thermal quenching in carbon-doped semi-insulating GaN

Jiahao Dong[1,2], Auditee Majumder Momo[3], Austin Fehr[1], Sanam SaeidNahaei[1], Pramod Reddy[3], Ronny Kirste[3], Zlatko Sitar[3], Ramón Collazo[3], Selim Elhadj[1,a]

[1)] Seurat Technologies, Wilmington, Massachusetts 01887, USA

[2)] Department of Materials Science and Engineering, Massachusetts Institute of Technology, Cambridge, Massachusetts 02139, USA

[3)] Department of Materials Science and Engineering, North Carolina State University, Raleigh, North Carolina 27695, USA

[a)] Corresponding author: elhadj@vt.edu

**Abstract:** Carbon-doped GaN is a promising material for sub-bandgap triggered optical switches. When incorporated in GaN, carbon introduces deep compensating centers that enable defect-mediated extrinsic photoconductivity. Here, we investigate the optical responsivity and switching kinetics of semi-insulating carbon-doped GaN actuated by sub-bandgap blue illumination. A high ON/OFF ratio exceeding $10^7$ is achieved under low-irradiance 405-nm excitation. Temperature-dependent transient measurements reveal that the photocurrent decay is thermally quenched above a crossover temperature of ~300 K. This behavior is attributed to hole-emission-assisted recombination. The extracted activation energies vary across samples; a commonly observed value of ~0.83 eV is attributed to the $C_N$ defect. Notably, when heating above the crossover temperature, thermal quenching accelerates the photocurrent decay by up to a factor of five, enabling significantly faster switching.

Carbon is one of the most important dopants in GaN, intentionally added or as an impurity. When incorporated during epitaxial growth, carbon can substitute on the nitrogen site, forming a deep acceptor $C_N$ with a (0/-) transition approximately 0.9 eV above the valence band.[1,2] This mid-gap state acts as an electron trap and compensates donors such as O or Si, rendering GaN semi-insulating.[3,4] As a result, carbon is widely employed to suppress buffer leakage in nitride-based electronic devices.[5–10]

Despite its extensive use in buffer layers in nitride electronics, carbon-doped GaN has remained largely unexplored as an active optoelectronic layer. Carbon incorporation introduces mid-gap states that, in principle, enable defect-mediated photoconductivity and sub-bandgap optical switching. Similar mechanisms have enabled sub-bandgap-triggered photoconductive semiconductor switches (PCSSs) in V-doped SiC, Ge-doped AlN, and N-doped diamond.[11–15] Yet, these material systems face practical limitations associated with epitaxial scalability and integration maturity. In contrast, C-doped GaN combines a mature epitaxial and processing ecosystem with well-established doping strategies, positioning it as a compelling but underexplored material for defect-engineered optical switching. More broadly, sub-bandgap response plays an important role in applications such as optically addressed spatial light modulators, photoelasticity and photoplasticity, and hysteretic electronic devices.[16–19]

In this work, we investigate sub-bandgap photoresponse and switching dynamics in semi-insulating carbon-doped GaN. We demonstrate high ON/OFF ratio exceeding $10^7$ at a low optical irradiance of 3.3 $mW/cm^2$ @ 405 nm. We systematically examine the temperature dependence of photoconductivity decay.

The decay transient is described by two components (slow and fast), both of which are thermally quenched above a crossover temperature of ~300 K. We attribute the thermal quenching to hole-emission-assisted recombination, which becomes the dominant carrier recombination pathway above the crossover temperature. We extracted activation energies of ~0.83 eV and 0.33 eV, depending on the sample and decay component; the commonly observed ~0.83 eV is likely associated with the $C_N$ (0/-) level. When heating above the crossover temperature, the photocurrent decay is accelerated by up to a factor of five, enabling significantly faster switching. These results highlight the important role of defect states in governing switching dynamics in GaN:C, particularly under elevated temperatures.

GaN:C samples grown by two different techniques were investigated, as illustrated in **Fig. 1**. **Fig. 1a** shows a 1-µm-thick C-doped GaN epilayer grown by metal organic chemical vapor deposition (MOCVD) on a c-plane ammonothermal GaN substrate under N-rich conditions. The C-doped layer was deposited on a 1-µm undoped GaN buffer layer. Detailed growth parameters are reported elsewhere.[20] In this sample, carbon concentration and combined Si and O donor concentration are both approximately $3.0\times10^{16}$ $cm^{-3}$. **Fig. 1b** shows a 400-µm-thick free-standing C-doped GaN single crystal grown by hydride vapor phase epitaxy (HVPE). In this sample, the carbon concentration is $2.7\times10^{17}$ $cm^{-3}$, while the combined Si and O donor concentration is approximately $1.0\times10^{17}$ $cm^{-3}$. In both samples [C] >= [Si] + [O], and thus the Fermi level is expected to be pinned around the deep acceptor level $C_N$ (0/-),[21] rendering the GaN semi-insulating as confirmed by electrical measurements.

For photoconductivity measurements and capacitance-voltage (C-V) profiling, vertical Schottky contacts were prepared on both samples: Ti/Al/Ti/Au was deposited on the bottom surface as ohmic contacts followed by rapid thermal annealing (RTA) at 950 °C for 30 s; Ni/Au (on the MOCVD GaN:C) or semitransparent Ni (on the HVPE GaN:C) contacts were deposited on the Ga-polar surface as Schottky contacts. On the HVPE GaN:C, the thin Ni layer has a thickness of 8 nm and 45% transmittance to 405 nm light.[22] The semitransparent electrode enables calculation of optical responsivity. During testing of the photoresponse, the illumination was incident on the top Schottky contacts.

For steady-state photoluminescence (PL) analysis, the GaN:C samples were excited with a 325 nm continuous wave HeCd laser with an irradiance of 0.1 $W/cm^2$. The PL emissions at room temperature were collected with a 0.75 m Acton SP2750 monochromator, and a PIXIS: 2KBUV cooled CCD camera. As shown in **Fig. 1c**, two dominant defect bands are observed in both samples: the yellow luminescence (YL) band with a maximum at ~2.2 eV, and the blue luminescence (BL) band with a maximum at ~2.9 eV. The YL band is typically associated with the $C_N$ (0/-) transition in C-doped GaN.[1,4,21,23] The BL band might be related to the $C_N$ (+/0) transition.[4,21] A detailed analysis will be presented in our future publication focusing on the PL results. We summarize the main defect transitions expected for both samples in **Fig. 1d** as suggested by the PL spectra and literature reports.

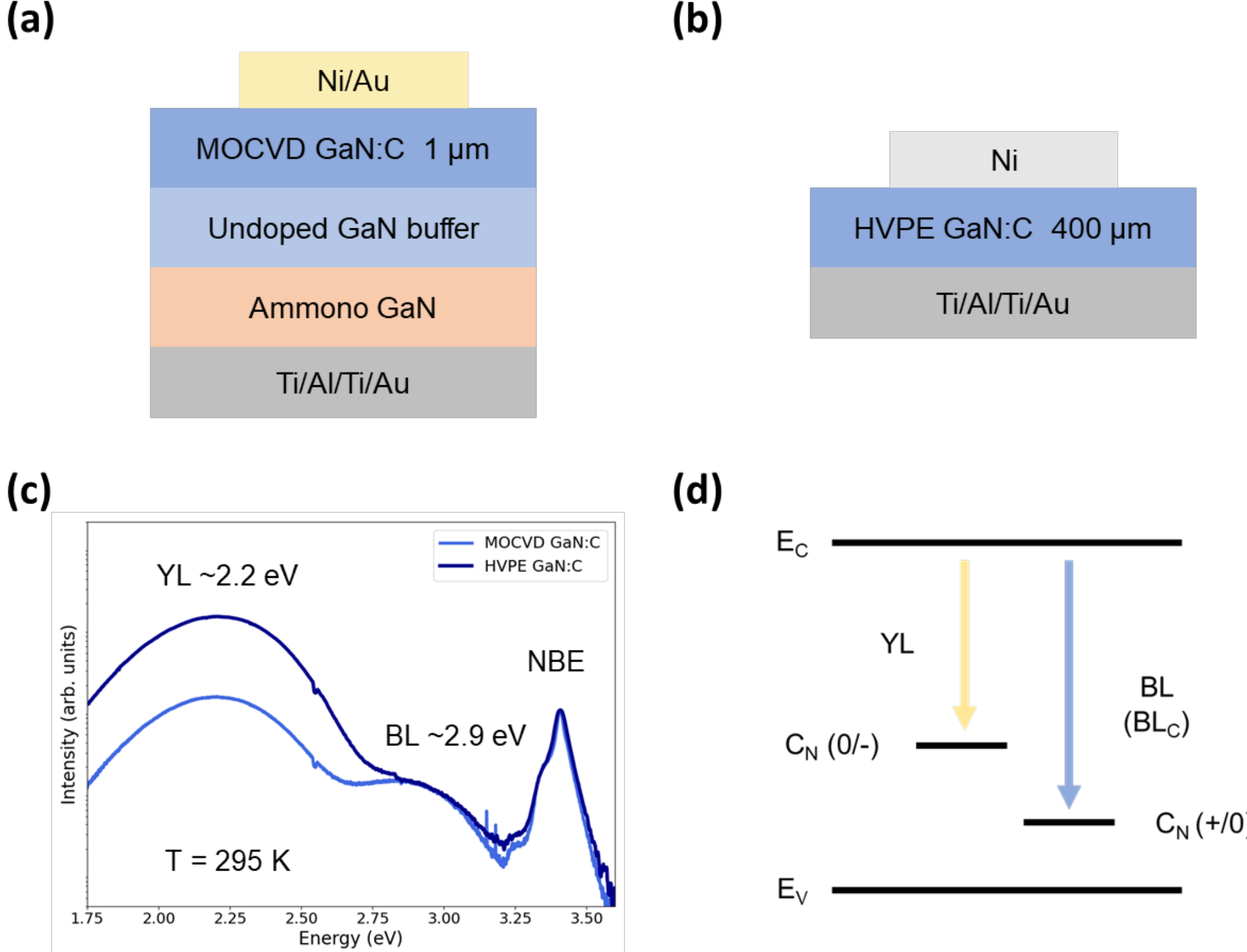

**FIG. 1:** Schematics of the sub-bandgap-triggered GaN:C optical switches, steady-state PL spectra and proposed main defect transitions. (a) GaN:C as well as undoped GaN buffer layer grown with MOCVD on an ammonothermal GaN substrate, with Ni/Au deposited on the top as Schottky contact. (b) HVPE GaN:C with an 8-nm semitransparent Ni layer deposited on the top as Schottky contact. On both samples, Ti/Al/Ti/Au was deposited on the bottom surface as ohmic contacts. (c) Steady-state PL spectra for the GaN:C in (a) and (b) measured at room temperature. YL, BL and NBE represent yellow luminescence, blue luminescence, and near-band-edge emission, respectively. (d) Proposed main defect transitions in the GaN:C as suggested by the PL spectra in (c) and by comparison with literature reports. The BL band might be related to the $BL_C$ band reported in [21].

**Fig. 2** describes I-V measurements of the GaN:C samples under dark and under illumination, respectively. The LED (Thorlabs SOLIS-405C) has a central wavelength of 405 nm (3.06 eV), below the GaN bandgap (3.40 eV). The dark current is below $10^{-7}$ mA/cm$^2$, confirming that the samples tested here are semi-insulating. Under light, for the MOCVD GaN:C, an ON/OFF ratio of $10^7$ is achieved at a forward bias of 1 V and a low irradiance of 3.3 mW/cm$^2$ @ 405 nm. For the HVPE GaN:C, an ON/OFF ratio of $10^9$ is achieved at the same bias and irradiance. The measured I-V curves of both samples show rectifying characteristics, as expected for Schottky diodes. The HVPE GaN:C shows a higher leakage current under reverse bias, as expected from a higher dislocation density in the HVPE-grown crystal. Further, we estimate the sub-bandgap responsivity of the HVPE GaN:C is 3.3 A/W @ 405 nm, by using a transmittance of 45% for the thin nickel layer.

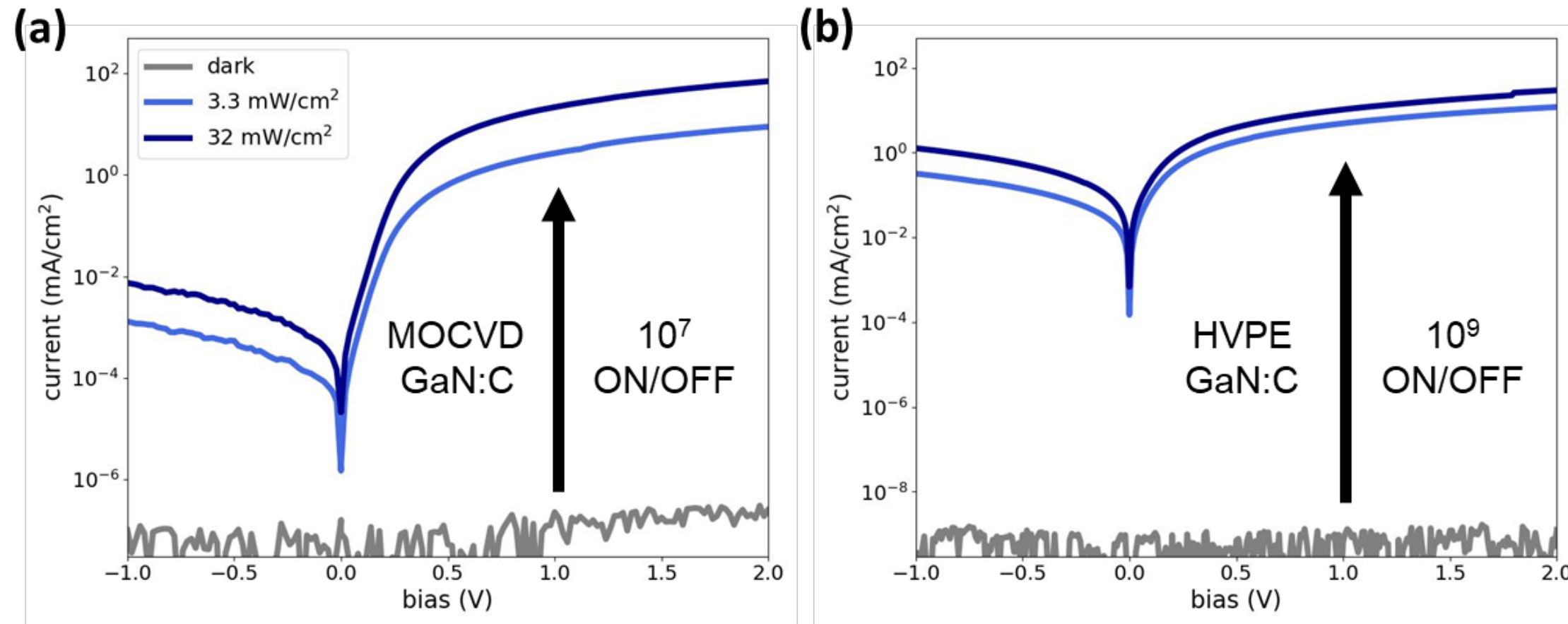

**FIG. 2:** Measured I-V curves under dark and light, respectively. Bias was applied on the top Schottky contact. Current measured from -1 to 2 V under dark (grey line) and 405-nm illumination at various irradiance (blue lines) for (a) the MOCVD GaN:C and (b) the HVPE GaN:C.

To estimate the carrier concentration induced by sub-bandgap light, we measured C-V @ 100 kHz with an LCR meter (HIOKI 3533). Under dark, due to very low concentration of background carriers, we observed constant capacitance versus forward or reverse bias for both samples, consistent with previous reports on GaN doped with [C] > [Si].[24] The sub-bandgap excitation can photoionize carbon-related deep levels and increase the *n*-type carrier concentration, enabling the extraction of the photocarrier concentration. Under 405-nm excitation of 32 mW/cm$^2$, we estimated a photocarrier concentration of $3.9\times10^{16}$ cm$^{-3}$ in the MOCVD GaN:C, and $2.5\times10^{17}$ cm$^{-3}$ in the HVPE GaN:C. These numbers are close to the nominal carbon doping concentration in both samples, suggesting that most of the carbon-related deep levels can be photoionized.

We performed transient photoconductivity measurements at various temperatures using a cryostat from Semetrol. The photocurrent transients enable the extraction of the current switching time—a key metric for optical switches—and the activation energies of the underlying defect transitions. Here, transient photocurrent decay was measured at a forward bias of 2 V on both GaN:C samples from 204 to 394 K, followed by a pulse of 405-nm excitation @ 220 mW/cm$^2$ for 5 s; the results are shown in the double-logarithmic plots in **Figs. 3a** and **3d**. To extract the decay rates, we use a two-exponential function to fit the decay transient $I(t)$, as shown by the dashed lines in **Figs. 3a** and **3d** (see supplementary material for tabulated fitting results at each temperature). The extracted decay rates $1/\tau_{\mathrm{slow}}$ and $/\tau_{\mathrm{fast}}$ are shown in the Arrhenius plots in **Figs. 3b** and **3e**. Notably, all the decay rates appear to feature two distinct thermally activated regimes. In the MOCVD GaN:C (**Fig. 3b**), both the slow and fast decay rates are almost temperature-independent at lower temperatures, and both show an activation energy of 0.83 eV at higher temperatures. The activation energies are extracted using the equation $1/\tau(T) \propto \left(1 + A\exp(-E_{\mathrm{a}}/kT)\right)/\tau_0$,[25] where $1/\tau(T)$ is the decay rate at temperature $T$, $E_{\mathrm{a}}$ the activation energy, and $A$ and $\tau_0$ are constants. The two-regime behavior of decay rates suggests a change of the dominant carrier recombination channel with increasing temperature. The crossover temperature $T_{\mathrm{C}}$ between the low and high temperature regimes is about 300 K. Similarly, for the HVPE GaN:C (**Fig. 3e**), both $1/\tau_{\mathrm{slow}}$ and $/\tau_{\mathrm{fast}}$ appear to be weakly, or temperature-independent at lower temperatures; however, at higher temperatures, they exhibit different activation energies: $E_{\mathrm{a,slow}} = 0.33$ eV for the slower decay component, and $E_{\mathrm{a,fast}} = 0.82$ eV for the faster decay component. The crossover temperature in the HVPE sample is similar to that in the MOCVD sample ($T_C$~300 K). Above 363 K the faster decay component $I_{\mathrm{fast}}$ completely disappears in the HVPE sample.

Importantly, the significant quenching of photocurrent decay above $T_C$ motivates us to use higher temperature operation to reduce turn-off time of the GaN:C optical switches. In **Figs. 3c** and **3f** we show 95% turn-off time (the time it takes for the photocurrent to decay to 5% of its steady state value) as a function of temperature for the two samples. For the MOCVD GaN:C, increasing temperature from 20 °C (*i.e.*, room temperature) to 70 °C reduces the turn-off time by 78% – from 9 to 2 ms, corresponding to an approximately fivefold acceleration of the photocurrent decay. For the HVPE GaN:C, heating from 20 °C to 70 °C reduces the turn-off time by 44% – from 9 to 5 ms.

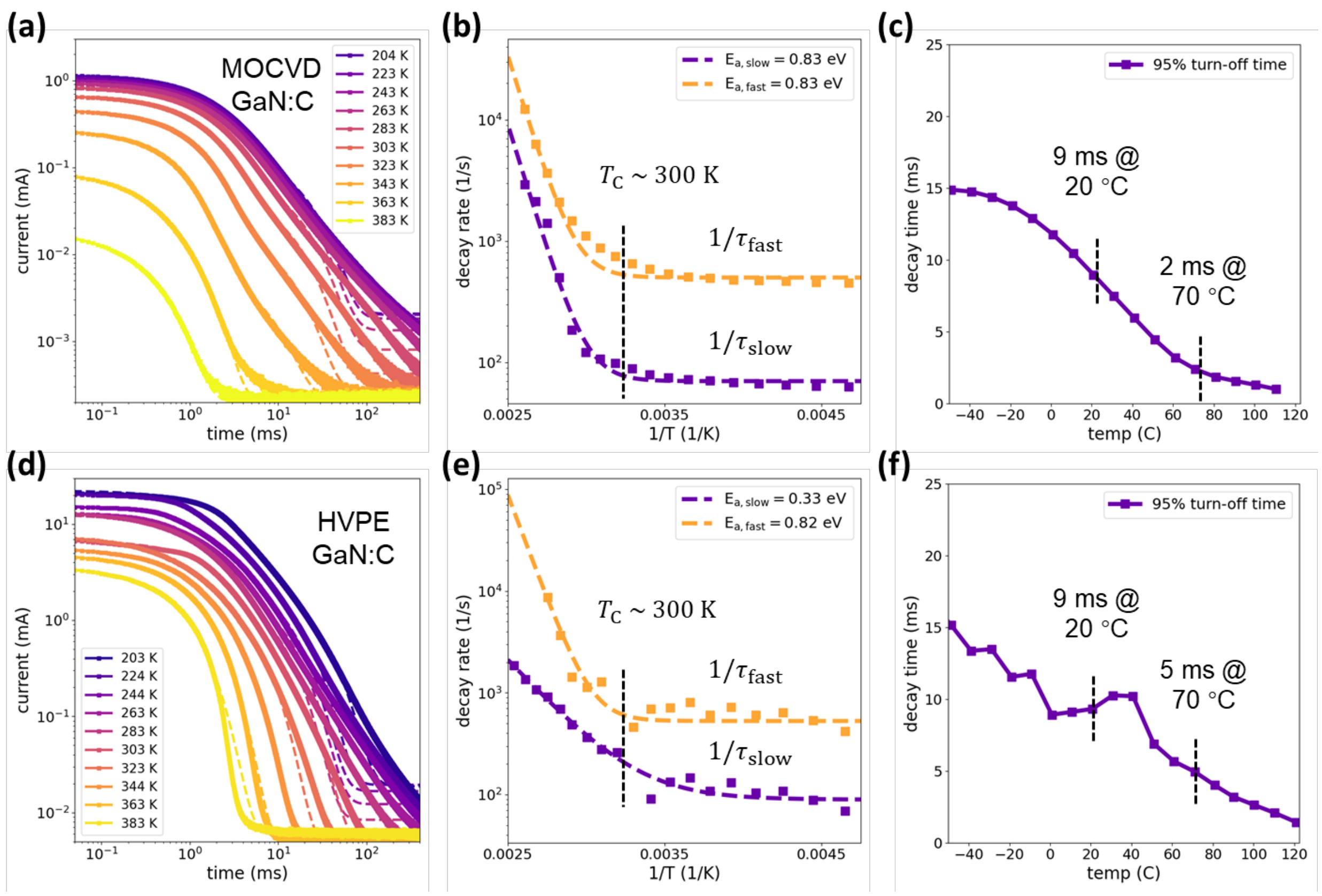

**FIG. 3:** Photocurrent transients, Arrhenius plots and turn-off times. (a) Photocurrent transients after light turning off. Dashed lines represent two-exponential fitting results. (b) Arrhenius plot for both the slow and fast decay rates. Dashed lines represent fitting with the equation $1/\tau(\mathrm{T}) \propto \left(1 + A\exp(-E_a/kT)\right)/\tau_0$, from which the activation energy $E_a$ is extracted. The crossover temperature $T_C$ between the low and high temperature regimes is indicated by a black dashed line. (c) 95% turn-off time. (a), (b) and (c) are for the MOCVD GaN:C, while (d), (e) and (f) are the same plots for the HVPE GaN:C.

To compare with previously reported sub-bandgap optical switches, we summarize the optical switching performance of the GaN:C samples tested in this work in **Table I**. GaN:C compares favorably in terms of responsivity, ON/OFF ratio, and turn-off time. The sub-bandgap responsivity of GaN:C is several orders of magnitude higher than that of diamond:N, and slightly lower than AlN:Ge. The strong optical response under sub-bandgap excitation allows effective triggering with visible light, eliminating the need for hazardous ultraviolet (UV) sources that can degrade UV-sensitive materials. Moreover, operation at elevated, yet still practical, temperatures significantly reduces turn-off time, enabling significantly faster optical switching. The turn-off time reported here corresponds to 95% decay. In future work, we look to identify and suppress the slower transient components beyond the 95% decay, enabling full utilization of the intrinsically high ON/OFF ratio (> $10^7$) of the GaN:C.

| Material | $\lambda_{exc}$ (nm) | R (A/W) | $I_{on}/I_{off}$ | Turn-off time | Reference |
|---|---|---|---|---|---|
| N-doped diamond | 532 | $7 \times 10^{-5}$ | - | - | Ref. [13] |
| Ge-doped AlN | 455 | 18 | $10^5$ | <100 ms | Ref. [12] |
| C-doped GaN [MOCVD] | 405 | - | $10^7$ | 2 ms @ 70 °C | This work |
| C-doped GaN [HVPE] | 405 | 3.3 | $10^9$ | 5 ms @ 70 °C | This work |

**TABLE I:** Performance comparison of the GaN:C studied in this work with other sub-bandgap-triggered WBG and UWBG materials. $\lambda_{exc}$ is the excitation wavelength for actuating the switch and R is the corresponding optical responsivity.

The thermal quenching of photocurrent decay above $T_C$ suggests a change of the dominant carrier recombination channel. To understand the switching dynamics in GaN:C, we model the carrier excitation and recombination pathways in the schematic band diagrams in **Fig. 4**. There is no band bending at the metal/GaN interface as the photoconductivity transients were measured under forward bias; thus, the carrier dynamics should not be affected by the space charge region at the Schottky junction. As shown in **Fig. 4a**, under sub-bandgap illumination, electrons trapped in carbon-related deep levels are excited to the conduction band, rendering GaN:C *n*-type. After light is turned off, decay of photocarriers may happen through two channels: direct electron recapture by the photoionized trap levels as shown in **Fig. 4b**; hole-emission-assisted e-h recombination as shown in **Fig. 4c**. The electron recapture by trapping centers is barrierless, and therefore the transition rate should be nearly independent of temperature. In contrast, the hole emission is thermally activated, with a barrier height of $E_t - E_V$, where $E_t$ is the trap level and $E_V$ the valence band level. At elevated temperatures, holes are emitted to the valence band from trap levels; these extra holes enable electron-hole recombination in all other channels (including radiative and nonradiative),[21] thus accelerating decay of *n*-type carriers and photocurrent transients.

The model in **Fig. 4** is consistent with the two-regime behavior of photocurrent decay rates: at lower temperature ($T < T_C$), the decay is dominated by electron recapture because of negligible concentration of thermally ionized holes, and consequently the decay rates show near-zero activation energies; at higher temperature ($T > T_C$), hole emission from deep traps is activated and hole-emission-assisted e-h recombination become the dominant decay channel. In the MOCVD GaN:C, both the slower and faster decay components show an activation energy of 0.83 eV, suggesting they may originate from the same defect level. One possible explanation is that these components arise from the same defect species located in the surface region and bulk region, respectively. However, further investigation is required to verify this hypothesis. In the HVPE GaN:C, we extracted an activation energy of 0.33 eV for the slower decay component, and 0.82 eV for the faster decay component. The ~0.83 eV activation energy observed across different samples is likely associated with the $C_N$ (0/-) transition, which has an energy level of around $E_t = E_v + 0.9$ eV.[1,21] The 0.33 eV activation is observed only in the HVPE sample. There are several defect candidates with an ionization energy in the range of 0.2-0.4 eV, including $C_N$ (+/0), and $Mg_{Ga}$ (0/-) and $Zn_{Ga}$ (0/-);[21,26,27] nonetheless, the ZPL of these transitions are reported higher than 3.06 eV, suggesting that they might not be excited by the 405-nm light source used in this study. Thus, the microscopic origin of the observed ~0.3 eV activation energy remains unresolved. Further, in the HVPE sample the 0.33 eV decay component exhibits a slower decay rate than the 0.82 eV component, which appears unusual. This suggests that the decay rate of the ~0.3 eV component may not be solely determined by the hole emission process, but may be affected by other processes such as carrier transport, e-h recombination dynamics and shallow impurity states. The microscopic origin of this behavior will be described in future work.

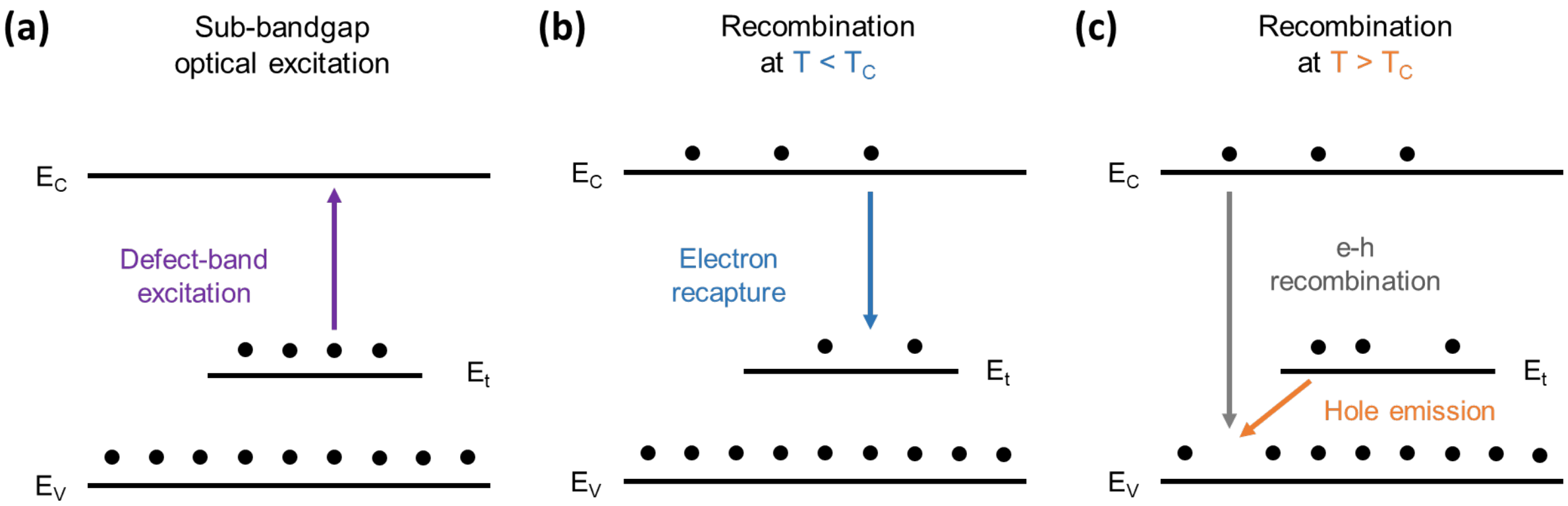

**FIG. 4:** Schematic band diagrams of the carrier excitation and recombination pathways in GaN:C. (a) Sub-bandgap illumination induces defect-to-band transitions, exciting electrons from the deep level $E_\mathrm{t}$ to the conduction band $E_\mathrm{C}$. (b) At $T < T_\mathrm{C}$, carrier recombination is dominated by direct electron recapture by the ionized deep level $E_\mathrm{t}$. (c) At $T > T_\mathrm{C}$, carrier recombination is dominated by hole-emission-assisted processes: holes are emitted from the ionized deep level $E_\mathrm{t}$ to the valence band $E_\mathrm{V}$, which then recombine with electrons in all other radiative and nonradiative channels (not drawn here).

The model shown in **Fig. 4** implies that the photocarrier decay rate is limited by the electron recapture rate $1/\tau_\mathrm{ec}$ at $T < T_\mathrm{C}$ and hole emission rate $1/\tau_\mathrm{he}$ at $T > T_\mathrm{C}$, respectively, which are given by the following equations:[19,28]

$$\frac{1}{\tau_\mathrm{ec}} = (N_\mathrm{t} - n_\mathrm{t}) v_\mathrm{rms} \sigma_\mathrm{n} \tag{1}$$

$$\frac{1}{\tau_\mathrm{he}} = N_\mathrm{V} v_\mathrm{rms} \sigma_\mathrm{p} \exp\left(-\frac{E_\mathrm{t} - E_\mathrm{V}}{k_\mathrm{B} T}\right) \tag{2}$$

where $N_\mathrm{t}$ and $n_\mathrm{t}$ are the total and occupied deep trap concentrations, $N_\mathrm{V}$ the valence band density of states, and $v_\mathrm{rms}$ the carrier root mean square velocity. $\sigma_\mathrm{n}$ and $\sigma_\mathrm{p}$ are the apparent electron and hole capture cross sections, respectively. Using Eq. (1) and Eq. (2), we extract $\sigma_\mathrm{n}$ and $\sigma_\mathrm{p}$ for the ~0.83 eV decay components that are likely associated with the $C_N$ (0/-) level; the results are summarized in **Table II**. Here, $N_\mathrm{t} - n_\mathrm{t}$ is estimated using the photocarrier concentration determined by C-V measurements. The extracted $\sigma_\mathrm{n}$ are very low, but are in agreement with those reported by Kato et al. ($3 \times 10^{-21}$ cm$^2$) and Polyakov et al. ($1.4 \times 10^{-22}$ cm$^2$),[29–31] suggesting that electron capture is intrinsically slow at the $C_N$ (0/-) level; however, the extracted $\sigma_\mathrm{p}$ appears one or two orders of magnitude higher than previously reported result ($\sim 10^{-14}$ cm$^2$), and the reason for this discrepancy is unclear. Although the decay's microscopic origin is complex, the pronounced acceleration of current switching with increasing temperature clearly demonstrates that thermal activation plays a key role, offering a practical path to faster photoconductive switching operation in carbon-doped GaN optoelectronic devices.

| Sample | MOCVD GaN:C | MOCVD GaN:C | HVPE GaN:C |
|---|---|---|---|
| Decay component | $1/\tau_{\text{slow}}$ | $1/\tau_{\text{fast}}$ | $1/\tau_{\text{fast}}$ |
| $\sigma_n$ | $1.6 \times 10^{-22}$ cm$^2$ | $1.2 \times 10^{-21}$ cm$^2$ | $1.7 \times 10^{-22}$ cm$^2$ |
| $\sigma_p$ | $7.1 \times 10^{-13}$ cm$^2$ | $2.6 \times 10^{-12}$ cm$^2$ | $5.1 \times 10^{-12}$ cm$^2$ |
| $E_t$ | $E_V + 0.83$ eV | $E_V + 0.83$ eV | $E_V + 0.82$ eV |
| Defect assignment | $C_N$ (0/-) | | |

**TABLE II:** Defect parameters extracted from photocurrent decays for the defect level that is likely associated with $C_N$ (0/-) in both GaN:C samples.

In conclusion, we demonstrate that strong sub-bandgap response can be achieved in semi-insulating GaN with moderate carbon doping concentration. The GaN:C shows a high ON/OFF ratio of over $10^7$ under low-irradiance blue illumination. Compared with previously reported sub-bandgap optical switches, the GaN:C presented here exhibit competitive optical responsivity, ON/OFF ratio, and switching speed. We examine the temperature dependence of photocurrent decay kinetics, and we found that thermal quenching happens above a crossover temperature $T_C$ (~300 K). We attribute this behavior to a change of the dominant carrier recombination pathway, *i.e.*, from electron recapture at $T < T_C$ to hole-emission-assisted recombination at $T > T_C$. Due to thermal quenching, the optical switching is accelerated by at most five times simply by heating the GaN:C from 20 °C to 70 °C. The ~0.83 eV activation energy observed across different samples is likely associated with the $C_N$ (0/-) level; the ~0.3 eV activation energy is observed only in the HVPE sample, and its defect assignment is unclear. While the microscopic origin of the underlying defect transitions requires further work to resolve, the significant acceleration of photoconductive switching with increasing temperature provides an important insight into achieving faster operation in GaN-based optoelectronic devices, and highlights the role of defect states in governing switching dynamics and material behavior at elevated temperatures.

## Supplementary Material

See supplementary material for two-exponential fitting results of the photocurrent decay at each temperature.

## References

[1] J.L. Lyons, A. Janotti, and C.G. Van De Walle, "Carbon impurities and the yellow luminescence in GaN," Appl. Phys. Lett. **97**(15), 152108 (2010).

[2] J.L. Lyons, E.R. Glaser, M.E. Zvanut, S. Paudel, M. Iwinska, T. Sochacki, and M. Bockowski, "Carbon complexes in highly C-doped GaN," Phys. Rev. B **104**(7), 075201 (2021).

[3] D.S. Green, U.K. Mishra, and J.S. Speck, "Carbon doping of GaN with CBr4 in radio-frequency plasma-assisted molecular beam epitaxy," J. Appl. Phys. **95**(12), 8456–8462 (2004).

[4] J.L. Lyons, A. Janotti, and C.G. Van De Walle, "Effects of carbon on the electrical and optical properties of InN, GaN, and AlN," Phys. Rev. B **89**(3), 035204 (2014).

[5] C. Poblenz, P. Waltereit, S. Rajan, S. Heikman, U.K. Mishra, and J.S. Speck, "Effect of carbon doping on buffer leakage in AlGaN/GaN high electron mobility transistors," J. Vac. Sci. Technol. B Microelectron. Nanometer Struct. Process. Meas. Phenom. **22**(3), 1145–1149 (2004).

[6] Z.-Q. Fang, B. Claflin, D.C. Look, D.S. Green, and R. Vetury, "Deep traps in AlGaN/GaN heterostructures studied by deep level transient spectroscopy: Effect of carbon concentration in GaN buffer layers," J. Appl. Phys. **108**(6), 063706 (2010).

[7] S.W. Kaun, M.H. Wong, J. Lu, U.K. Mishra, and J.S. Speck, "Reduction of carbon proximity effects by including AlGaN back barriers in HEMTs on free-standing GaN," Electron. Lett. **49**(14), 893–895 (2013).

[8] A. Fariza, A. Lesnik, J. Bläsing, M.P. Hoffmann, F. Hörich, P. Veit, H. Witte, A. Dadgar, and A. Strittmatter, "On reduction of current leakage in GaN by carbon-doping," Appl. Phys. Lett. **109**(21), 212102 (2016).

[9] Ö. Danielsson, X. Li, L. Ojamäe, E. Janzén, H. Pedersen, and U. Forsberg, "A model for carbon incorporation from trimethyl gallium in chemical vapor deposition of gallium nitride," J. Mater. Chem. C **4**(4), 863–871 (2016).

[10] W. Cao, C. Song, H. Liao, N. Yang, R. Wang, G. Tang, and H. Ji, "Numerical simulation analysis of carbon defects in the buffer on vertical leakage and breakdown of GaN on silicon epitaxial layers," Sci. Rep. **13**(1), 14820 (2023).

[11] T. He, T. Shu, H. Yang, M. Yi, F. Liu, J. Yao, L. Wang, and T. Xun, "Effect of Donor–Acceptor Compensation on Transient Performance of Vanadium-Doped SiC Photoconductive Switches Using 532-nm Laser," IEEE Trans. Electron Devices **71**(7), 4275–4282 (2024).

[12] J. Dong, and R. Jaramillo, "A Junction Photoconductive Semiconductor Switch (J-PCSS) in AlN with Sub-Band Gap Responsivity and Accelerated Turn-Off Speed," IEEE Electron Device Lett. **46**(6), 916–919 (2025).

[13] D.L. Hall, L.F. Voss, P. Grivickas, M. Bora, A.M. Conway, P. Scajev, and V. Grivickas, "Photoconductive Switch with High Sub-Bandgap Responsivity in Nitrogen-Doped Diamond," IEEE Electron Device Lett. **41**(7), 1070–1073 (2020).

[14] E. Majda-Zdancewicz, M. Suproniuk, M. Pawłowski, and M. Wierzbowski, "Current state of photoconductive semiconductor switch engineering," Opto-Electron. Rev. **26**(2), 92–102 (2018).

[15] V. Meyers, L. Voss, J.D. Flicker, L.G. Rodriguez, H.P. Hjalmarson, J. Lehr, N. Gonzalez, G. Pickrell, S. Ghandiparsi, and R. Kaplar, "Photoconductive Semiconductor Switches: Materials, Physics, and Applications," Appl. Sci. **15**(2), 645 (2025).

[16] S. Elhadj, Z. Davidson, and Y. Sargol, "A light-driven light valve for metal additive manufacturing," in *Smart Mater. Opto-Electron. Appl.*, edited by I. Rendina, L. Petti, D. Sagnelli, and G. Nenna, (SPIE, Prague, Czech Republic, 2023), p. 3.

[17] B. Chatterjee, S. Ghandiparsi, M.S. Gottlieb, Q. Shao, C.D. Frye, S. Harrison, and L. Voss, "Wide bandgap photoconductor (SiC:V)-based optically addressed light valve for high fluence operation," Appl. Opt. **64**(10), 2324 (2025).

[18] J. Dong, Y. Li, Y. Zhou, A. Schwartzman, H. Xu, B. Azhar, J. Bennett, J. Li, and R. Jaramillo, “Giant and Controllable Photoplasticity and Photoelasticity in Compound Semiconductors,” Phys. Rev. Lett. **129**(6), 065501 (2022).
[19] J. Dong, and R. Jaramillo, “Modeling defect-level switching for nonlinear and hysteretic electronic devices,” J. Appl. Phys. **135**(22), 224501 (2024).
[20] S. Rathkanthiwar, P. Bagheri, D. Khachariya, S. Mita, S. Pavlidis, P. Reddy, R. Kirste, J. Tweedie, Z. Sitar, and R. Collazo, “Point-defect management in homoepitaxially grown Si-doped GaN by MOCVD for vertical power devices,” Appl. Phys. Express **15**(5), 051003 (2022).
[21] M.A. Reshchikov, M. Vorobiov, D.O. Demchenko, Ü. Özgür, H. Morkoç, A. Lesnik, M.P. Hoffmann, F. Hörich, A. Dadgar, and A. Strittmatter, “Two charge states of the C N acceptor in GaN: Evidence from photoluminescence,” Phys. Rev. B **98**(12), 125207 (2018).
[22] M.A. Nasiri, A. Seijas-Da Silva, J.F. Serrano Claumarchirant, C.M. Gómez, G. Abellán, A. Cantarero, and J. Canet-Ferrer, “Ultrathin Transparent Nickel Electrodes for Thermoelectric Applications,” Adv. Mater. Interfaces **11**(5), 2300705 (2024).
[23] S.G. Christenson, W. Xie, Y.Y. Sun, and S.B. Zhang, “Carbon as a source for yellow luminescence in GaN: Isolated CN defect or its complexes,” J. Appl. Phys. **118**(13), 135708 (2015).
[24] C.H. Seager, A.F. Wright, J. Yu, and W. Götz, “Role of carbon in GaN,” J. Appl. Phys. **92**(11), 6553–6560 (2002).
[25] M.A. Reshchikov, D.O. Demchenko, A. Usikov, H. Helava, and Yu. Makarov, “Carbon defects as sources of the green and yellow luminescence bands in undoped GaN,” Phys. Rev. B **90**(23), 235203 (2014).
[26] M.A. Reshchikov, P. Ghimire, and D.O. Demchenko, “Magnesium acceptor in gallium nitride. I. Photoluminescence from Mg-doped GaN,” Phys. Rev. B **97**(20), 205204 (2018).
[27] D.O. Demchenko, and M.A. Reshchikov, “Blue luminescence and Zn acceptor in GaN,” Phys. Rev. B **88**(11), 115204 (2013).
[28] K. Decock, P. Zabierowski, and M. Burgelman, “Modeling metastabilities in chalcopyrite-based thin film solar cells,” J. Appl. Phys. **111**(4), 043703 (2012).
[29] A.Y. Polyakov, N.B. Smirnov, E.B. Yakimov, S.A. Tarelkin, A.V. Turutin, I.V. Shemerov, S.J. Pearton, K.-B. Bae, and I.-H. Lee, “Deep traps determining the non-radiative lifetime and defect band yellow luminescence in n-GaN,” J. Alloys Compd. **686**, 1044–1052 (2016).
[30] M. Kato, T. Asada, T. Maeda, K. Ito, K. Tomita, T. Narita, and T. Kachi, “Contribution of the carbon-originated hole trap to slow decays of photoluminescence and photoconductivity in homoepitaxial n-type GaN layers,” J. Appl. Phys. **129**(11), 115701 (2021).
[31] M. Kato, T. Maeda, K. Ito, K. Tomita, T. Narita, and T. Kachi, “Relationship of carbon concentration and slow decays of photoluminescence in homoepitaxial n-type GaN layers,” Jpn. J. Appl. Phys. **61**(7), 078004 (2022).